\documentclass[aps,prl,twocolumn,groupedaddress,a4paper,floatfix,showpacs]{revtex4}
\usepackage{graphicx}

\begin{document}
\title{Non-linear unbalanced Bessel beams:\\ Stationary conical waves supported by nonlinear losses}
\author{Miguel A. Porras$^1$, Alberto Parola$^2$, Daniele Faccio$^2$, Audrius Dubietis$^3$
        and Paolo Di Trapani$^2$}
\affiliation{$^1$Departamento de F\'{\i}sica Aplicada, Universidad Polit\'ecnica de Madrid, Rios
             Rosas 21, E-28003 Madrid, Spain \\
             $^2$INFM and Department of Physics, University of Insubria, Via Valleggio 11, IT-22100
             Como, Italy \\
             $^3$Department of Quantum Electronics, Vilnius University, Sauletekio 9, LT-10222 Vilnius,
             Lithuania
             }
\begin{abstract}
Nonlinear losses accompanying self-focusing substantially impacts the dynamic balance of diffraction and
nonlinearity, permitting the existence of localized and stationary solutions of the 2D+1 nonlinear
Schr\"{o}dinger equation which are stable against radial collapse. These are featured by linear, conical
tails that continually refill the nonlinear, central spot. An experiment shows that the discovered solution
behaves as strong attractor for the self-focusing dynamics in Kerr media.
\end{abstract}


\maketitle

One of the main goals of modern nonlinear wave physics is the achievement of wave localization,
stationarity and stability. While in a one-dimensional geometry (e.g., in optical fibers), nonlinearity
suitably balances linear wave dispersion, leading to the soliton regime, in the multidimensional case,
nonlinearity drives waves either to collapse or instability. In self-focusing of optical beams, for
instance, many stabilizing mechanisms, such as Kerr saturation, plasma-induced defocusing, or stimulated
Raman scattering, have been explored, and are being the subject of intense debate, mainly in the context of
light filamentation in air or condensed matter \cite{COUAIRON}.
These mechanisms, however, are either intrinsically lossy, or, due to the huge intensities involved, are
accompanied by losses, which lead ultimately to the termination of any soliton regime. Similar pictures can
be traced in all phenomena commonly discussed in the context of the nonlinear Schr\"odinger equation
(NLSE), as Bose-Einstein condensates (BEC) or Langmuir waves in plasma \cite{SULEM}. Nonlinear losses (NLL)
arise in BEC from two- and three-body inelastic recombination, and as the natural mechanism for energy
dissipation in Langmuir turbulence \cite{RUSSIAN}.

The question then arises of whether any stationary and localized (SL) wave propagation is possible in the
presence of NLL. The response, as shown in this Letter, is affirmative. These SL waves cannot be
ascribed to the class of solitary waves, but are instead {\em nonlinear conical waves} (as the non-linear X
waves \cite{TRAPANI}) of {\em dissipative} type, whose stationarity is sustained by a continuous refilling
of the nonlinearly absorbed central spot with the energy supplied by linear, conical tails. These waves are
not only robust against NLL, but find their stabilizing mechanism against perturbations in NLL themselves.

Among the linear conical waves \cite{SALO}, the simplest one is the monochromatic Bessel beam (BB)
\cite{DURNIN}, made of a superposition of plane waves whose wave vectors are evenly distributed over the
surface of a cone, resulting in a nondiffracting transversal Bessel profile. Despite the ideal nature of
BBs (they carry infinite power), they not only have revealed to be a paradigm for understanding wave
phenomena, but also have found applications as diverse as in frequency conversion, or in atom
trapping and alignment \cite{NATURE}. Of particular interest for us is the finding \cite{CHAVEZ} that the
BB is describable in terms of the interference of two conical Hankel beams \cite{SALO}, carrying equal
amounts of energy towards and outwards the beam axis, and yielding no net transversal energy flux in
the BB.

What we demonstrate here is that a superposition of inward and outward Hankel beams with unequal
amplitudes, i.e., an ``unbalanced" Bessel beam (UBB), describes the only possible asymptotic form of the
SL, nonsingular solutions of the 2D+1 NLSE with NLL. For this reason, we call the solution ``non-linear
UBB" (NL-UBB). We then show that NLL-UBB solutions do exist and are stable against radial perturbations, in
the important case of Kerr nonlinearity with NLL. Unbalancing, that manifests as a reduced
visibility of the radial Bessel oscillations, creates the required inward radial energy flux from the
conical tails of the beam to refill the nonlinearly absorbed central spot, whose transversal
pattern depends on the specific nonlinear phase (nondissipative) terms included in the NLSE. Contrary to
linear conical waves, achievement of stationarity by refilling imposes generally a lower bound to the cone
angle of the UBB.

In a self-focusing experiment in water, we demonstrate the self-generation of a NL-UBB from a
Gaussian wave-packet, which evidences that the NL-UBB acts as a strong attractor in the self-focusing
dynamics, and hence a significant role of NLL in light filamentation.




To start with, we consider the 2D+1 NLSE
\begin{equation}\label{NLSE}
\partial_z A = \frac{i}{2k}\nabla^2_\perp A + i\omega\frac{n_2}{c}|A|^2 A -
\frac{\beta^{(K)}}{2}|A|^{2K-2}A ,
\end{equation}
for the propagation of a light beam $E=A\exp(-i\omega t + ik z)$ of frequency $\omega$ in a Kerr
medium (other nonlinear phase terms could be included as well) with NLL. In (\ref{NLSE}),
$\nabla_\perp\equiv(\partial_{x},\partial_{y})$, $k=n\omega/c$ is the propagation constant (with $n$ the
refraction index and $c$ the speed of light in vacuum), $n_2$ is the nonlinear refraction index, and
$\beta^{(K)}>0$ ($K=2,3,\dots$) is the multi-photon absorption coefficient. For the real amplitude and
phase [$A=a\exp(i\varphi)$, $a>0$], (\ref{NLSE}) yields
\begin{eqnarray}
\partial_z a^2  &=& - \frac{1}{k}\nabla_\perp\cdot(a^2\nabla_\perp \varphi) - \beta^{(K)}a^{2K}
\label{INT}, \\
\partial_z\varphi &=&\frac{1}{2k}\left[\frac{\nabla^2_\perp a}{a}-
(\nabla_\perp\varphi)^2\right] + \frac{\omega n_2}{c}a^2. \label{PHASE}
\end{eqnarray}
Stationarity of the intensity profile ($\partial_z a^2 = 0$) requires, from (\ref{INT}), $\varphi=\phi(x,y)
+ g(z)$. Then (\ref{PHASE}) imposes the linear dependence $g(z)= - \delta z$, where $\delta$ is a constant
wave vector shift. Equations (\ref{INT}) and (\ref{PHASE}) then lead, for cylindrical beams, to the
eigenvalue problem
\begin{eqnarray}\label{PHSTA}
a^{\prime\prime} \! +\! \frac{a'}{r} \!+\! 2k\delta\, a - \!(\phi')^2
a \!+\! 2 \frac{k^2 n_2}{n}a^3&=&0 ,\\
-\frac{1}{k}2\pi r \phi' a^2 = \beta^{(K)}2\pi \int_0^r dr r a^{2K} &\equiv & N_r, \label{ENERGY}
\end{eqnarray}
[prime signs stand for $d/dr$, with $r\equiv(x^2+y^2)^{1/2}$], with boundary conditions $a(0)\equiv a_0
> 0$, $a'(0)=0$, $\phi'(0) = 0$, and the requirement that $a(r)\rightarrow 0$ as $r\rightarrow
\infty$ for localization. Eq. (\ref{ENERGY}) establishes that in a SL beam, the power (per unit propagation
length) entering into a disk of radius $r$ must equal the power lost $N_r$ within it. In absence of NLL
($\beta^{(K)}=0$), this condition demands plane phase fronts ($\phi'= 0$) and no radial energy flux. This
is the case of the sech-type Townes profile in Kerr media \cite{CHIAO}, associated to wave vector shift
$\delta<0$, and of weakly localized Bessel beams in linear \cite{DURNIN} or Kerr media \cite{JOHANNINSSON},
with infinite power and wave vector shift $\delta>0$.

With NLL, instead, the conditions of refilling (\ref{ENERGY}) and localization [$a(r)\rightarrow 0$ as
$r\rightarrow \infty$] require an inward radial power [lhs of (\ref{ENERGY})] that monotonically increases
[rhs of (\ref{ENERGY})] with $r$ up to reach, at infinity, a constant value equal to the total NLL,
$N_\infty$, assumed they are finite. Stationarity with NLL is thus supported by the continuous refilling of
the more strongly absorbed inner part of the beam with the energy coming from its outer part. Phase fronts
cannot be plane, since $\phi'\rightarrow -kN_\infty/2\pi r a^2$ as $r\rightarrow \infty$. As for the
amplitude, the asymptotic value of $\phi'$ and the change $a(r)=b(r)/\sqrt{r}$ in (\ref{PHSTA}), lead, when
retaining only the slowest decaying contributions, to the Newton-like equation $b^{\prime\prime}=-2k\delta
b + k^2N_\infty^2/4\pi^2b^3$, that represents the ``motion" of a particle in the potential $V(b)=k\delta
b^2 + k^2N_\infty^2/8\pi^2b^2$. Since bounded trajectories $b(r)$ [leading then to $a\rightarrow 0$] under
this potential can exist only for strictly positive $\delta$, we conclude that {\em SL waves in media with
NLL can only have positive wave vector shift,} $\delta>0$. The solution of the Newton equation then yields
the asymptotic behavior $a(r)=\{[c_1 + c_2\cos(2\sqrt{2k\delta}\,r +c_3)]/r\}^{1/2}$, with $c_1>0$,
$c_1^2-c_2^2=kN_\infty^2/8\pi^2\delta$, that represents radial oscillations of contrast $C=|c_2|/c_1$ about
an equilibrium point that approaches zero as $1/\sqrt{r}$. SL beams in media with NLL carry then infinite
power, and have superluminal phase velocity ($\delta>0$). These asymptotic features are more meaningfully
expressed in terms of the UBB
\begin{equation}\label{UBB}
A\!\simeq\!\frac{a_0}{2}\left[\alpha_{\mbox{\small out}}H_0^{(1)}(\sqrt{2k\delta}r) \!+\!
\alpha_{\mbox{\small in}}H_0^{(2)}(\sqrt{2k\delta} r) \right]e^{-i\delta z}\! ,
\end{equation}
formed by two nondiffracting Hankel beams of the first and second kind \cite{SALO}, of same cone angle
$\theta=\sqrt{2\delta/k}$, but different weights, $\alpha_{\mbox{\small out}}$ and
$\alpha_{\mbox{\small in}}$, that must be related by
\begin{equation}\label{UNBALANCE}
a_0^2(|\alpha_{\mbox{\small in}}|^2 - |\alpha_{\mbox{\small out}}|^2)/k=N_\infty .
\end{equation}
The balanced superposition of Hankel beams (e.g., $\alpha_{\mbox{\small out}}=\alpha_{\mbox{\small in}}=1$)
just gives the original, nondiffracting Bessel beam $a_0J_0(\sqrt{2k\delta}\,r)\exp(-i\delta z)$, with no
net radial energy flux, and with oscillations of maximum contrast $C=1$. Instead, unbalancing creates an
inward radial power that is manifested in a lowering of the contrast $C=|c_2|/c_1 = 2|\alpha_{\mbox{\small
in}}||\alpha_{\mbox{\small out}}|/(|\alpha_{\mbox{\small in}}|^2+|\alpha_{\mbox{\small out}}|^2)$ of the
Bessel oscillations, reaching $C=0$ (no oscillations) in a pure Hankel beam.

\begin{figure}
\begin{center}
\includegraphics[width=8.6cm]{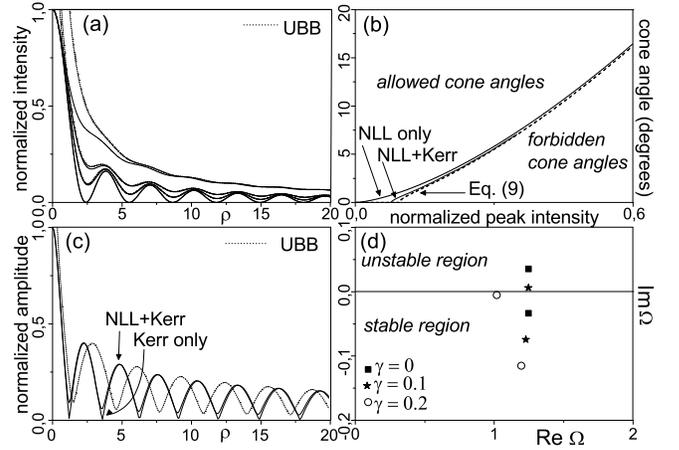}
\end{center}
\caption{\label{nokerr} (a) For pure NLL ($K=4$), radial profiles $a^2/a_0^2$ of SL beams with decreasing
$\delta/\beta^{(K)}I_0^{K-1}=\infty, 0.33, 0.25, 0.20$ (from lower to higher ones), and their asymptotic
UBB. Normalized radial coordinate is $\rho=\sqrt{2k\delta} r$ (b) Allowed cone angles
$\theta=\sqrt{2\delta/k}$ versus peak intensity $I_0$ normalized to the
intensity $(k/\beta^{(K)})^{1/(K-1)}$ in the cases of pure NLL, and NLL+Kerr (numerically calculated and
from Eq. \ref{deltakerr}). (c) For NLL+Kerr, amplitude profile $a/a_0$ for
$\delta/\beta^{(K)}I_0^{K-1}=0.24$ and $kn_2I_0/n\delta=6.07$, and its asymptotic UBB. The pure Kerr case
is also shown. (d) Eigenvalue spectrum of the perturbations with $m=0$ to the SL solutions with Kerr
nonlinearity $g=1$ and NLL ($K=4$) $\gamma=0, 0.1, 0.2$.}
\end{figure}

We stress that this analysis holds irrespective of the nonlinear phase terms in the NLSE (Kerr
nonlinearity, Kerr saturation, \dots), since they rely on the only assumption of finite NLL. We can
therefore state that {\em the conical UBB represents the only possible asymptotic form of SL waves in
nonlinear media when the effects of NLL are taken into consideration.} Note that the case of linear losses
is excluded, since $N_\infty=\infty$ for $K=1$. It should be also clear that the actual existence of a SL
solution of the NLSE, and the characteristics of its linear asymptotic UBB (cone angle,
$\alpha_{\mbox{\small in}}$ and $\alpha_{\mbox{\small out}}$), depends on the particular nonlinear phase
terms in the NLSE.

We first solved numerically (\ref{PHSTA}) and (\ref{ENERGY}) without the Kerr term, to appreciate the
effects of pure NLL [Fig. \ref{nokerr}(a)], and found that SL solutions [$a(r)\rightarrow 0$] exist indeed
with any peak intensity $I_0=a_0^2$ and wave vector shifts
\begin{equation}\label{deltanokerr}
\delta > g_K \beta^{(K)}I_0^{K-1} \, ,
\end{equation}
where $g_K =$ 1.67, 0.27, 0.19, 0.16 $\dots$ for $K=$ 2, 3 $\dots$ As seen in Fig. \ref{nokerr}(a), shortly
away from the central peak, the radial profile becomes undistinguishable from that of the UBB of same
$\delta$, matched NLL $N_\infty$ and contrast $C$. Moreover, {\em pure NLL creates a gap in the allowed UBB
cone angles $\theta=\sqrt{2\delta/k}$ that increases with intensity} [Fig. \ref{nokerr}(b)], and that
becomes significant at intensities comparable to the characteristic intensity $(k/\beta^{(K)})^{1/(K-1)}$.
Intuitively, this gap originates from the fact that the SL profiles widen and delocalize [Fig.
\ref{nokerr}(a)] as the cone angle diminishes. This causes the total NLL, $N_\infty$, to increase, a
situation that cannot be sustained down to the limit of zero cone angle ($\delta\rightarrow 0$), which
would not allow for any radial energy flux.

The extension to include nonlinear phase terms in the NLSE can be readily understood from the case of pure
NLL. Figure \ref{nokerr}(b) shows the permitted cones angles in the case of focusing Kerr nonlinearity
($n_2>0$), obtained from numerical integration of (\ref{PHSTA}) and (\ref{ENERGY}) with $K=4$ (similar
results hold for other values of $K$). The modification of the allowed region of cone angles can be
attributed to the nonlinear phase shift at the central spot. In fact, assuming that the existence of a
localized solution is now determined by the effective wave vector shift $\delta_{\mbox{\small eff}}=\delta
+ \delta_{\mbox{\small nl}}$, with $\delta_{\mbox{\small nl}}= kn_2I_0/n$ for Kerr nonlinearity, we replace
$\delta$ with $\delta_{\mbox{\small eff}}$ in Eq. (\ref{deltanokerr}), to obtain
\begin{equation}\label{deltakerr}
\delta > \mbox{max}\left\{ g_K\beta^{(K)}I_0^{K-1}\, -\, kn_2I_0/n\, , \,\, 0\right\}
\end{equation}
(where we set to zero negative values) as an accurate expression for the allowed linear wave vector shifts
in Kerr media [see Fig. \ref{nokerr}(b) for the associated cone angles $\theta=\sqrt{2\delta/k}$]. Similar
expressions as (\ref{deltakerr}) can be obtained for other nonlinear phase effects.

When NLL dominate over Kerr nonlinearity [right part of Fig. \ref{nokerr}(b), or $g_K\beta^{(K)}
I_0^{K-1}\gg kn_2I_0/n_2$] the SL profiles (not shown) do not substantially differ from the case of pure
NLL. For the Kerr-dominated case [left part of Fig. \ref{nokerr}(b), or $g_K\beta^{(K)} I_0^{K-1}\ll
kn_2I_0/n_2$], Fig. \ref{nokerr}(c) shows a representative SL profile. The central peak and inner rings are
nearly identical to the Kerr-compressed, Bessel-like beam in lossless Kerr media \cite{JOHANNINSSON}, as
seen in Fig. \ref{nokerr}(c), though the small NLL leads to a slight contrast reduction. The central peak
can be approached by the Bessel beam $a_0J_0(\sqrt{2k\delta_{\mbox{\small eff}}}\,r)$ \cite{JOHANNINSSON},
and hence its width by $1/\sqrt{2k\delta_{\mbox{\small eff}}}$. Outer rings gradually shrink up to become
phased with the asymptotic UBB of wave vector shift $\delta$ [Fig. \ref{nokerr}(c)]. Note that, as opposed
to the pure NLL regime, the beam does not widen indefinitely as the cone angle diminishes down to its lower
bound ($\theta\rightarrow 0^{+}$ in the Kerr-dominated case), but is limited to a maximum beam width
$1/\sqrt{2k\delta_{\mbox{\small eff}}}= \sqrt{n/2k^2n_2I_0}$. This fact entails, contrary to the pure NLL
regime, a limitation to $N_\infty$ as $\theta\rightarrow 0^{+}$, explaining why the unbalance mechanism for
replenishment can support stationarity at arbitrarily small, but positive cone angles.

We also studied the stability of the NL-UBB solutions against perturbations. With the dimensionless
quantities $\rho=\sqrt{2k\delta}\,r$, $\xi=\delta z$ and ${\cal A} =A/a_0$, we rewrite (\ref{NLSE}) as
\begin{equation}\label{NNLSE}
\partial_\xi {\cal A} = i\nabla^2_\rho {\cal A} + i g |{\cal A}|^2{\cal A} -\gamma |{\cal
A}|^{2K-2}{\cal A}
\end{equation}
where $\nabla^2_\rho= \partial^2_\rho + (1/\rho)\partial_\rho$, $g=\omega n_2 a_0^2/c\delta$ and
$\gamma=\beta^{(K)}a_0^{2K-2}/2\delta$. Following a standard Bogoliubov-deGennes procedure \cite{SKRYABIN},
we introduce a perturbed solution
\begin{equation}
{\cal A} = {\cal A}_0(\rho,\xi) + \left[ u(\rho)e^{-i\Omega \xi + i m\theta} - v^{*}(\rho)e^{i\Omega^{*}\xi
- im\theta}\right]e^{-i\xi} ,\nonumber
\end{equation}
where ${\cal A}_0(\rho,\xi)= a(\rho)\exp[i\phi(\rho)]\exp(-i\xi)$ is a SL solution of (\ref{NNLSE}),
$\theta$ is the polar angle, and $m=0,1\dots$, into (\ref{NNLSE}), to obtain, upon linearization, the
(non-self-adjoint) eigenvalue problem
\begin{eqnarray}
\!\Omega u &\!\!=\!\!& H u \!-\! i K\gamma a^{2K\!-\!2} u \!+\! g a^2\! e^{2i\phi}v \!+\! i
(K\!-\!1)\gamma a^{2K-2}\! e^{2i\phi} v \nonumber \\
\!-\!\Omega v &\!\!=\!\!& H v \!+\! iK\gamma a^{2K\!-\!2} v \!+\! g a^2\! e^{\!-2i\phi}
u\!-\!i(K\!\!-\!\!1)\gamma a^{2K\!-\!2}\! e^{\!-2i\phi} u ,\nonumber \\ \label{NSA}
\end{eqnarray}
where $H= -\nabla^2_\rho + m^2/\rho^2 - 1 - 2g a^2$, with boundary conditions $u(\rho)\rightarrow 0$,
$v(\rho)\rightarrow 0$ as $\rho\rightarrow\infty$. The existence of an eigenvalue $\Omega$ with
$\mbox{Im}\Omega>0$ would render unstable the SL solution ${\cal A}_0$. To obtain numerically the
eigenvalues, we transform (\ref{NSA}) into an algebraical eigenvalue problem by introducing a mesh of small
size $s$ in the range $[\rho=0,\rho=L]$ with large $L$, writing the differential operators as finite
differences, and imposing the boundary conditions $u(L)=v(L)=0$. The numerical diagonalization of the
$2N\times 2N$ matrix ($N$ being the number of mesh points) of the eigenvalue problem provides a set of $2N$
eigenvalues. The behavior of the continuous distribution of eigenvalues of (\ref{NSA}) is inferred by
extrapolating the results to the limit $s\rightarrow 0$ and $L\rightarrow \infty$. We investigated
increasing $N$ up to 4000 ($L=400$, $s=0.1$), as limited by the accessible memory of our computational
facility.

Figure \ref{nokerr}(d) shows a typical eigenvalue spectrum for radial perturbations ($m=0$) to three
SL solutions with same Kerr nonlinearity ($g=1$) and increasing NLL. Eigenvalues with
$\mbox{Im}\Omega=0$ are not shown for clarity. Clearly, {\em the effect of NLL is to decrease the imaginary
part of the eigenvalues, driving the system towards stability}: The Bessel-like solutions in pure Kerr
media (squares) are unstable. At NLL strength $\gamma=0.1$ (stars), the positive imaginary parts are
strongly reduced, and at $\gamma=0.2$ (circles) no signs of instability are present. We performed the same
analysis for $m=1, 2, \dots$ For dipolar perturbations ($m=1$) no instability emerges neither in the pure
Kerr nor in the Kerr + NLL model. In contrast, both systems turned out to be unstable for quadrupolar
and higher-order perturbations, a result that can be related to the fact these perturbations involve
modulation far from the central spot, where intensity is weak and so NLL cannot play its stabilizing
role.



%
\begin{figure}
\begin{center}
\includegraphics[width=8.6cm]{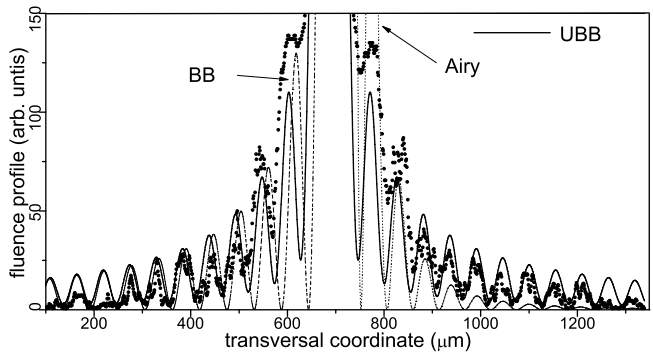}
\end{center}
\caption{\label{experiment} Measured fluence profile (dots), fitted UBB, and half Bessel (left) and Airy
(right) profiles.}
\end{figure}

The experiment that we present demonstrates that the NL-UBB stationary solution indeed acts as strong
attractor for the transient dynamics of light beam self-focusing in (weakly dispersive) Kerr media. Recently
\cite{DUBIETIS} we have shown that light filaments in water do not behave as
soliton-like beams, but they spread after being clipped by an aperture and reconstruct themselves
after being blocked by a stopper, as expected for genuine conical waves \cite{HERMAN}. Here we concentrate
our attention in the beam-periphery structure in order to demonstrate that filaments are
indeed conical and, more precisely, NL-UBB waves. To this end we modified the diagnostic by adopting
professional digital photo-camera (Canon-EOS D30), which has the unique advantage of permitting strong
local saturation (by the central spike) without any blooming effect. The experiment was done by launching a
spatially-filtered, collimated, 200 fs, 0.1 mm, 1.5$\,\mu$J, 527 nm Gaussian wave packet into a 31 mm
water-filled cuvette. By using an imaging spectrograph we verified that, at the specified pump energy, no
relevant spectral broadening occurs, the self-phase modulation occurring in the sole spatial domain. Figure
\ref{experiment} shows the measured fluence profile at the output facet of the non-linear sample, and a
fitted UBB profile. BB and Airy profiles (only one half, for clarity) are also shown for comparison. One can
appreciate: (i) The Bessel-like decay ($\sim 1/r$) of the measured profile in a
fairly vast region of the beam, which distinguishes it sharply from any (Airy-type) aperture-diffraction
pattern (with faster decay $\sim 1/r^2$). (ii) The accurate fitting of the UBB radial modulations to those
of the measured profile in the full recorded area, which even reproduces the increasing frequency of the
modulations (respect to a BB) towards the beam center, attributable to Kerr self-focusing. (iii) The
reduction in modulation contrast (compared to a BB), which is a signature of unbalance between inward and
outward conical power flows.

In conclusion, we have reported on the existence, characteristics, stability and experimental
relevance of non-linear unbalanced-Bessel-beams (NL-UBB), i.e., the stationary and localized solutions
of the 2D+1 NLSE in the presence of NLL. NL-UBB are asymptotically linear conical waves, carrying a net
inward power flux that compensate for the NLL. We have shown that the UBB asymptotics is the sole
compatible with NLL, no matter which are the specific non-linear phase terms in the NLSE. Therefore NL-UBB
are possible solutions also of nonintegrable NLSEs as well as of the Gross-Pitaevskii equation for
Bose-Einstein condensates. Owing to the spatiotemporal analogy, the results directly apply to pulse
propagation in planar wave guides with anomalous dispersion. More generally, similar waves should exist also
in the case of 3D+1 NLSE with NLL, describing full spatiotemporal localization in bulk media
with NLL. The unique property of long-range stationarity in the presence of energy transfer to matter
(or to other waves) makes NL-UBB ideal for several applications including deep-field non-linear microscopy,
laser micro machining, laser-writing of channel wave guides, charged-particle acceleration, creation of long
and stable plasma channels in atmosphere and, of course, energy transfer between different types of waves
(e.g., optical $\rightarrow$ X, or optical $\rightarrow$ teraHerz).


%

\end{document}